\documentclass[11pt,letter]{article}
\usepackage[letterpaper]{geometry}
\usepackage{times}
\usepackage{amsfonts}
\usepackage{stmaryrd}
\usepackage{bbm}
\usepackage{timestamp}
\usepackage{smallsec}
\usepackage{pgf}
\usepackage{tikz}
\usetikzlibrary{arrows,automata}
\usepackage[latin1]{inputenc}

\usepackage{amssymb}
\usepackage{amsmath}
\usepackage{amscd}
\usepackage{latexsym}
\usepackage{paralist}

\newtheorem{theorem}{Theorem}[section]
\newtheorem{dfn}[theorem]{Definition}      %
\newenvironment{definition}[1]{\begin{dfn}[{\textbf{#1}}]\rm}{ $\lrcorner$\end{dfn}}
\newtheorem{rmrk}[theorem]{Remark}
\newenvironment{remark}{\begin{rmrk}\rm}{\hspace{\stretch{1}}$\lrcorner$\end{rmrk}}

\def\squarebox#1{\hbox to #1{\hfill\vbox to #1{\vfill}}}
\newcommand{\qed}{\hspace*{\fill}
	    \vbox{\hrule\hbox{\vrule\squarebox{.667em}\vrule}\hrule}\smallskip}
\newenvironment{proof}{\begin{trivlist}
\item[\hspace{\labelsep}{\bf\noindent Proof: }]
}{\qed\end{trivlist}}

\newcommand{\mSection}[1]{\section{#1}}
\newcommand{\mSubSection}[1]{\subsection{#1}}

\newcommand{\Tau}{\Upsilon}

\newcommand{\BXMPL}{\begin{example}}
\newcommand{\EXMPL}{\end{example}}
\newcommand{\BPF}{\begin{proof}}
\newcommand{\EPF}{\end{proof}}
\newcommand{\BPR}{\begin{proposition}}
\newcommand{\EPR}{\end{proposition}}
\newcommand{\BCJR}{\begin{conjecture}}
\newcommand{\ECJR}{\end{conjecture}}
\newcommand{\BT}{\begin{theorem}}
\newcommand{\ET}{\end{theorem}}
\newcommand{\BCR}{\begin{corollary}}
\newcommand{\ECR}{\end{corollary}}
\newcommand{\BRM}{\begin{remark}}
\newcommand{\ERM}{\end{remark}}
\newcommand{\BCM}{\begin{claim}}
\newcommand{\ECM}{\end{claim}}

\newcommand{\BE}{\begin{enumerate}}
\newcommand{\EE}{\end{enumerate}}
\newcommand{\BI}{\begin{itemize}}
\newcommand{\EI}{\end{itemize}}
\newcommand{\bc}{\begin{center}}
\newcommand{\ec}{\end{center}}

\newcommand{\zug}[1]{\la{#1}\ra}

\newcommand{\buchi}{B$\ddot{\textrm{u}}$chi}

\newcommand{\U}{{\cal U}}

\newcommand{\A}{{\cal A}}
\newcommand{\B}{{\cal B}}

\renewcommand{\S}{{\cal S}}

\newcommand{\sG}{{\hspace{-.05mm}_{\cal G}\hspace{-.3mm}}}

\newcommand{\atstate}{\textit{node}}

\newcommand{\stam}[1]{}
\newcommand{\dual}{\textit{dual}}

\newcommand{\la}{{\langle}}
\newcommand{\ra}{{\rangle}}

\newcommand{\APT}{APT}

\newcommand{\nashg}{\textsc{nash$_<$}}
\newcommand{\nashge}{\textsc{nash}}

\newcommand{\dominatingg}{\textsc{ds$_<$}}

\newcommand{\dominatingge}{\textsc{ds}}

\newcommand{\spege}{\textsc{spe}}
\newcommand{\speg}{\textsc{spe$_<$}}

\newcommand{\play}{p}
\newcommand{\Plays}{\mathcal{P}}
\newcommand{\Actions}{\Sigma}
\newcommand{\action}{\sigma}

\newcommand{\ESL}{ESL}
\newcommand{\StrategyVars}{\mathbb{Z}}

\newcommand{\HistoryVars}{\mathbb{H}}
\newcommand{\z}{z}
\newcommand{\h}{{h}}

\newcommand{\Next}{\bigcirc}
\newcommand{\Until}{\,\mathcal{U}\,}
\newcommand{\Free}{{\textit{free}}}

\newcommand{\ph}{q_{\textit{his}}}
\newcommand{\pf}{q_{\textit{fut}}}
\newcommand{\pbot}{q_{\textit{rej}}}
\newcommand{\ptop}{q_{\textit{acc}}}

\newcommand{\actiona}{{\textit{$a_1$}}}
\newcommand{\actionA}{{\textit{$a_2$}}}
\newcommand{\actionb}{{\textit{$b_1$}}}
\newcommand{\actionB}{{\textit{$b_2$}}}
\newcommand{\actionc}{{\textit{$c_1$}}}
\newcommand{\actionC}{{\textit{$c_2$}}}

\newcommand{\wactiona}{{\textit{$\ \ \ a_1$}}}
\newcommand{\wactionA}{{\textit{$a_2\ \ \ $}}}
\newcommand{\wactionb}{{\textit{$b_1\ \ $}}}
\newcommand{\wactionB}{{\textit{$\ \ \ b_2$}}}
\newcommand{\wactionc}{{\textit{$\ \ \ c_1$}}}
\newcommand{\wactionC}{{\textit{$c_2\ \ \ $}}}

\newcommand{\fsystem}{\f_0}

\newcommand{\eexists}{\exists}
\newcommand{\fforall}{\forall}

\newcommand{\X}{\mathbb{X}}

\newcommand{\sema}[1]{[ \hspace{-0.05cm} [ {#1} ] \hspace{-0.05cm} ]}

\newcommand{\nmodels}{{\begin{picture}(14,12)\put(3,0){$\models$}\put(5.5,-1){$\slash$}\end{picture}}}

\newcommand{\outcome}{\textit{outcome}}
\newcommand{\payoff}{\textit{payoff}}

\newcommand{\myin}{\,\!\in\!\,}
\newcommand{\myneq}{{\textsf{$\neq$}}}%
\newcommand{\tightin}{{\textsf{$\in$}}}

\newcommand{\tuple}[1]{\langle {#1} \rangle}
\newcommand{\comment}[1]{}

\newcommand{\draftcomment}[1]{\texttt{\textbf{CHECK}: {#1}}}

\newcommand{\ltl}{LTL}

\newcommand{\agentsset}{I}
\newcommand{\agentssetwo}{I_{-0}}
\newcommand{\player}[1]{\textit{Agent {#1}}}

\newcommand{\f}{\varphi}
\newcommand{\g}{{\psi}}

\newcommand{\true}{{\sc\bf True}}
\newcommand{\false}{{\sc\bf False}}

\newcommand{\lalways}[1]{{\it always~#1}}
\newcommand{\leventually}[1]{{\it eventually~#1}}

\newcommand{\G}{\mathcal{G}}

\newcounter{sidecommentcounter}

\renewcommand{\L}{{\cal L}}

\newcommand{\join}{\lor}
\newcommand{\meet}{\land}

\begin{document}

\title{Rational Synthesis}
\author{Dana Fisman\\Hebrew University and IBM Haifa \and Orna Kupferman\\Hebrew University \and Yoad Lustig\\Rice University}
\date{}
\maketitle
\thispagestyle{empty}


\begin{abstract}
{\em Synthesis} is the automated construction of a system from its
specification. The system has to satisfy its specification in all possible environments.
Modern systems often interact with other systems, or agents.
Many times these agents have objectives of their own, other than to fail
the system. Thus, it makes sense to model system environments not as
hostile, but as composed of {\em rational agents}; i.e., agents that act to
achieve their own objectives.

We introduce the problem of synthesis in the context of rational
agents ({\em rational synthesis}, for short). The input
consists of a temporal-logic formula specifying the system and temporal-logic
formulas specifying the objectives of the agents. The output is an
implementation $T$ of the system and a profile of strategies,
suggesting a behavior for each of the agents.
The output should satisfy two conditions. First, the
composition of $T$ with the strategy profile should satisfy the
specification. Second, the strategy profile should be an equilibria in the sense that, in view of  their objectives, agents
have no incentive to deviate from the strategies assigned
to them.
We solve the rational-synthesis problem for various definitions of
equilibria studied in game theory.
We also
consider the multi-valued case in which the objectives of the system
and the agents are still temporal logic formulas, but involve payoffs from a
finite lattice.
\end{abstract}


\mSection{Introduction}\label{intro}
{\em Synthesis} is the automated construction of a system from its
specification. The basic idea is simple and appealing:
instead of developing a system and verifying that it adheres to its
specification, we would like to have an automated procedure that, given
a specification, constructs a system that is correct by construction.
The first formulation of synthesis goes back to Church \cite{Chu63};
the modern approach to synthesis was initiated by Pnueli and Rosner,
who introduced LTL (linear temporal logic) synthesis \cite{PR89a}.
In LTL synthesis, the specification is given in LTL and the output is a reactive system modeled by a finite-state transducer.
Much of today's research in formal verification is aimed at increasing
the practicality of automated synthesis, and it addresses challenges
like simplification of synthesis algorithms \cite{KV05a},
compositionality and modularity \cite{KPV06,LV09}, extensions of the
basic setting to richer ones (c.f., synthesis of distributed systems,
concurrent systems, and on-line algorithms \cite{AKL09,AAE04,KV01,MW05}), and extensions of the underline techniques to further
applications (c.f. automated control and repair \cite{JGB05,RW89}).

In synthesis, there is a distinction between system
outputs, controlled by the system, and system inputs, controlled by
the environment.
A system should be able to cope with all values of the input signals, while
setting the output signals  to desired values \cite{PR89a}.
Therefore, the quantification structure on input and output signals
is different. Input signals are universally quantified while output
signals are existentially quantified.

Modern systems often interact with other systems.
For example, the clients interacting with a server are
by themselves distinct entities (which we call agents) and are many
times implemented by systems.
In the traditional approach to synthesis, the way in which the
environment is composed of its underlying agents is abstracted.
In
particular, the agents can be seen as if their only objective is to
conspire to fail the system. Hence the term ``hostile environment" that
is traditionally used in the context of synthesis.
In real life, however, many times agents have goals of their own,
other than to fail the system. The approach taken in the field of
algorithmic game theory \cite{NRTV07} is to assume that agents interacting with a
computational system are {\em rational}, i.e., agents act to achieve
their own goals.
Assuming agents rationality is a restriction on the agents behavior
and is therefore equivalent to restricting the universal
quantification on the environment.
Thus, the following question arises: can system synthesizers capitalize on the
rationality and goals of agents interacting with the system?

Consider for example a peer-to-peer network with only two agents.
Each agent is interested in downloading infinitely often, but has no
incentive to upload. In order, however, for one
agent to download, the other agent must upload.
More formally, for each ${i\in\{0,1\}}$, Agent~$i$ controls the bits $u_i$
(``Agent~$i$ tries to upload'') and $d_i$ (``Agent~$i$ tries to download'').
The objective of Agent~$i$ is $\lalways{\leventually{(d_i\land u_{1-i})}}$.
Assume that we are asked to synthesize the protocol for Agent~0.
It is not hard to see that the objective of Agent~0 depends on
his input signal, implying he cannot ensure his objective in the
traditional synthesis sense. On the other hand, suppose that Agent~0, who is aware of the objective of Agent~1, declares and follows the
following {\sc tit for tat} strategy: I will upload at
the first time step, and from that point onward I will reciprocate the
actions of Agent~1.
Formally, this amounts to initially setting $u_0$ to
\true\  and for every time $k>0$, setting $u_0$ at time $k$ to equal $u_1$
at time $k-1$.
It is not hard to see that, against this
strategy, Agent~1 can only ensure his objective by satisfying Agent~0
objective as well. Thus, assuming Agent~1 acts rationally, Agent~0 can
ensure his objective.

The example above demonstrates that a synthesizer can capitalize on
the rationality of the agents that constitute its environment.
When synthesizing a protocol for rational agents, we still have no
control on their actions. We would like, however, to
generate a strategy for each agent (a \emph{strategy profile}) such
that once the strategy profile is given to the agents, then a rational
agent would have no incentive to deviate from the strategy suggested
to him and would follow it.
Such a strategy profile is called in game theory a {\em solution} to the game.
Accordingly, the {\em rational synthesis\/} problem gets as input temporal-logic  formulas  specifying the objective $\f_0$ of the system and the
objectives $\f_1,\ldots,\f_n$ of the agents that constitute the environment.
The desired output is a system and a strategy profile for the agents such that the following hold.
First, if all agents adhere to their strategies, then the result of the
interaction of the system and the agents satisfies $\f_0$.
Second, once the system is in place, and the agent are playing a game
among themselves, the strategy profile
is a solution to this game.\footnote{For a formal definition of \emph{rational synthesis}, see Definition~\ref{def:rationalsynt}.}

\stam{
The example above demonstrates that a synthesizer can capitalize on
the rationality of agents involved. On the other hand, the example
omits intricacies of the interaction between rational agents.
Assume, for example, that one is not asked to
synthesize Agent 0 in the peer-to-peer network, but is rather asked to synthesize a
protocol to be used by the two symmetric agents. Since the agents are
not directly controlled by the protocol synthesizer, one cannot assume
that the agents would adhere to the published protocol.
Thus,
when synthesizing a protocol for rational agents, we would like to
generate a strategy for each agent (a \emph{strategy profile}) such
that once the strategy profile is given to the agents, every rational
agent follows the strategy assigned to him.
Such a strategy profile is called in game theory a {\em solution} to the game.
}

%
A well known solution concept is
{\em Nash equilibrium}~\cite{Nas50}. 
 A strategy profile is in Nash equilibrium
if no agent has an incentive to deviate from his assigned strategy,
provided that the other agents adhere to the strategies assigned to them.
For example, if the {\sc tit for tat} strategy for Agent 0 is suggested to both agents, then the pair of
strategies is a Nash equilibrium. Indeed, for all $i \in \{0,1\}$, if Agent $i$  assumes that Agent $1-i$ adheres to his strategy, then by following the strategy, Agent $i$ knows that his objective would be satisfied, and he has no incentive to deviate from it.
The stability of a Nash equilibrium depends on the players assumption
that the other players adhere to the strategy. In some cases this is
a reasonable assumption. Consider, for example, a standard protocol published by some known authority such
as IEEE. When a programmer writes a program implementing the standard,
he tends to assume that his program
is going to interact with other programs that implement the same standard.
If the published standard is a Nash equilibrium, then there is no incentive
to write a program that diverts from the standard.
Game theory suggests several {\em solution concepts}, all capturing
the idea that the participating agents have no incentive to deviate
from the protocol (or strategy) assigned to them.
We consider three well-studied solution concepts~\cite{NRTV07}:
dominant-strategies solution, Nash equilibrium, and subgame-perfect
Nash equilibrium.

\stam{
In our first peer-to-peer example, the synthesis was of a system whose environment
consists of a single rational agent. In the second example, the
synthesis was of a protocol published for all agents, but there was no system to synthesize.
In the general case, one is asked to synthesize both a system
(e.g., a server) and strategies for the agents (e.g., clients).
Thus, the {\em rational synthesis\/} problem gets as input temporal logic  formulas  specifying the objective $\f_0$ of the system and the
objectives $\f_1,\ldots,\f_n$ of the agents that constitute the environment.
The desired output is a system and a strategy profile for the agents such that the following hold.
First, if all agents adhere to their strategies, then the result of the
interaction of the system and the agents satisfies $\f_0$.
Second, once the system is in place, and the agent are playing a game
among themselves, the strategy profile
is a solution to this game.\footnote{For a formal definition of \emph{rational synthesis}, see Definition~\ref{def:rationalsynt}.}
}

An important facet in the task of a rational synthesizer is to synthesize a system
such that once it is in place, the game played by the agents
has a solution with a favorable outcome. {\em Mechanism design},
studied in game theory and economy \cite{NR99,NRTV07},  is the study
of designing a game whose outcome (assuming
players rationality) achieves some goal. Rational synthesis can be
viewed as a variant of mechanism design in which the game is induced
by the objective of the system, and the objectives of both the
system and the agents refer to their on-going interaction and
are specified by temporal-logic  formulas.

Having defined rational synthesis, we turn to solve it.
In \cite{CHP07}, the authors introduced {\em strategy logic} -- an
extension of temporal logic with first order quantification over strategies.
The rich structure of strategy logic enables it to specify
properties like the existence of a Nash-equilibrium. While
\cite{CHP07} does not consider the synthesis problem, the technique
suggested there can be used in order to solve the rational-synthesis
problem for Nash equilibrium and dominant strategies.
Strategy logic, however, is not sufficiently expressive in order to specify
subgame-perfect-Nash equilibrium ~\cite{Sel75} which, as advocated in~\cite{Umm06} (see also Section 3), is the most suited for infinite multiplayer games  --- those induced by rational synthesis.
The weakness of strategy logic is its inability to quantify over game histories.
We extend strategy logic with history variables, and show that the extended logic is sufficiently expressive to express rational synthesis for the three solution concepts we study. Technically, adding history variables to strategy logic results in a {\em memoryful logic}~\cite{KV06a}, in which temporal logic  formulas have to be evaluated not along  paths that start at the present, but along paths that start at the root and go through the present.

Classical applications of game theory consider games with real-valued
payoffs. For example, agents may bid on goods or grade candidates.
In the peer-to-peer network example, one may want to refer to the
amount of data uploaded by each agent, or one may
want to add the possibility of pricing downloads.
The full quantitative setting is undecidable already in the
context of model checking~\cite{CCHKM05}. Yet, several special cases for
which the problem is decidable have been studied~\cite{CDH08}. We can
distinguish between cases in which decidability is achieved by
restricting the type of systems~\cite{CCHKM05}, and cases in which it is
achieved by restricting the domain of values~\cite{GC03}.
We solve the quantitative rational synthesis problem for the case the
domain of values
is a finite distributive De Morgan lattice.
The lattice setting is a good starting
point to the quantitative setting. First, lattices have been
successfully handled for easier problems, and in particular,
multi-valued synthesis~\cite{KL07,KL07b}. In addition,
lattices are sufficiently rich to express
interesting quantitative properties.
This is sometime immediate (for example,
in the peer-to-peer network, one can refer to the different
attributions of the communication channels, giving rise to
the lattice of the subsets of the attributions), and sometimes
thanks to the fact that real values can often be abstracted
to finite linear orders. From a technical point of view, our contribution here is a solution of a latticed game in which the value of the game cannot be obtained by joining values obtained by different strategies, which is unacceptable in synthesis.

\mSubSection{Related Work}\label{sec:discusssion}

Already early work on synthesis has realized that working with a hostile environment is often too restrictive. The way to address this point, however, has been by adding assumptions on the environment, which can be part of the specification (c.f., \cite{CHJ08}).
The first to consider the game-theoretic approach to dealing with rationality of the environment in the context of LTL synthesis were Chatteerjee and Henzinger~\cite{CH07}. 
The setting in~\cite{CH07}, however, is quite restricted; it considers exactly three players, where the third player is a fair scheduler, and the notion of \emph{secure equilibria}~\cite{CHJ06}. Secure equilibria, introduced in~\cite{CHJ06}, is a Nash equilibria in which each
of the two players prefers outcomes in which only his objective is achieved over outcomes in which both objectives are achieved, which he still prefers over outcomes in which his objective is not achieved.  It is not clear how this notion can be extended to multiplayer games, and to the distinction we make here between controllable agents that induce the game (the system) and rational agents (the environment). Also, the set of solution concepts we consider is richer.

Ummels~\cite{Umm06} was the first to consider subgame perfect
equilibria in the context of infinite multiplayer games. The
setting there is of turn-based games and the solution goes
via a reduction to 2-player games. Here, we consider concurrent
games and therefore cannot use such a reduction.
Another difference is that \cite{Umm06} considers parity winning
conditions whereas we use LTL objectives.
In addition, the fact that the input to the rational synthesis problem does not include a game makes the memoryful nature of subgame perfect equilibria more challenging, as we cannot easily reduce the LTL formulas to memoryless parity games.


To the best of our knowledge, we are the first to handle the
multi-valued setting. As we show, while the lattice case
is decidable, its handling required a nontrivial extension of
both the Boolean setting and the algorithms known for solving
latticed games \cite{KL07b}.

%



\mSection{Preliminaries}\label{subsec:games}

We consider \emph{infinite concurrent multiplayer games} (in short, \emph{games}) defined as follows. A \emph{game arena} is a tuple $\G= \la V,v_0, \agentsset,(\Actions_i)_{i\in\agentsset}, (\Gamma_i)_{i\in\agentsset},\delta \ra$, where $V$ is a set of nodes,
$v_0$ is an initial node,
$I$ is a set of players, and for $i \in I$, the set
$\Actions_i$ is the set of actions of Player~$i$ and $\Gamma_i:V \rightarrow 2^{\Actions_i}$ specifies the actions that Player~$i$ can take at each node. Let $\agentsset = \{1,\ldots,n\}$. Then, the transition relation
$\delta:V \times \Actions_1 \times \cdots \times \Actions_n \rightarrow V$ is a deterministic function mapping the current node and the current choices of the agents to the successor node.  The transition function may be restricted to its relevant domain. Thus, $\delta(v,\action_1,\ldots,\action_n)$ is defined for $v  \in V$ and  $\zug{\action_1,\ldots,\action_n} \in$ $\Gamma_1(v) \times \cdots \times \Gamma_n(v)$.

A {\em position} in the game is a tuple $\la v, \action_1,\action_2,\ldots ,\action_n\ra$ with ${v\myin V}$ and $\action_i\in \Gamma_i(v)$ for every $i\in I$. Thus, a position describes a state along with possible choices of actions for the players in this state.
Consider a sequence $\play=\play_0 \cdot \play_1 \cdot  \play_2 \cdots$ of positions.
For $k \geq 0$, we use $\atstate(\play_k)$ to denote the state component of $\play_k$, and use $\play_k[i]$, for $i\in\agentsset$, to denote the action of Player~$i$ in $\play_k$. The notations extend to $p$ in the straightforward way. Thus, $\atstate(\play)$ is the projection of $p$ on the first component.
We say that $\play$ is a \emph{play} if the transitions between positions is consistent with $\delta$. Formally, $\play$ is a \emph{play
starting at node $v$} if $\atstate(\play_0)=v$ and for all $k \geq 0$, we have $\atstate(\play_{k+1})=\delta(\play_k)$.
We use $\Plays_\G$ (or simply $\Plays$ when $\G$ is clear from the context) to denote all possible plays of $\G$.
%

Note that at every node ${v\myin V}$, each player $i$ chooses an action  ${\action_i\myin\Gamma_i(v)}$ simultaneously and independently of the other players. The game then proceeds to the successor node $\delta(v,\action_1,\ldots,\action_n)$.
A \emph{strategy} for Player~$i$ is a function ${\pi_i: V^+ \mapsto \Actions_i}$ that maps histories of the game to an action suggested to Player~$i$. The suggestion has to be consistent with $\Gamma_i$. Thus, for every ${v_0v_1 \cdots v_k \in V^+}$, we have ${\pi_i(v_0v_1 \cdots v_k)\myin\Gamma_i(v_k)}$. Let $\Pi_i$ denote the set of possible strategies for Player~$i$.
For a set of players $\agentsset = \{1,\ldots,n\}$, a \emph{strategy profile} is
 a tuple of strategies ${\la \pi_1,  \pi_2, \ldots, \pi_n\ra \in \Pi_1 \times \Pi_2 \times \cdots \times \Pi_n}$. We denote the strategy profile by $(\pi_i)_{i\in\agentsset}$ (or simply $\pi$, when $I$ is clear from the context). 
   We say that $\play$ is an \emph{outcome} of the profile $\pi$ if
for all ${k \geq 0}$ and ${i\in\agentsset}$, we have
${\play_{k}[i]=\pi_i(\atstate(\play_{0})\cdot \atstate(\play_{1})\cdots \atstate(\play_{k}))}$. Thus, $\play$ is an outcome of $\pi$ if all the players adhere to their strategies in $\pi$. Note that since $\delta$ is deterministic, $\pi$ fixes a single play from each state of the game. Given a profile $\pi$ we denote by $\outcome(\pi)^\sG$ (or simply $\outcome(\pi)$) the one play in $\G$ that is the outcome of $\pi$ when starting in $v_0$.  Given a strategy profile $\pi$ and a nonempty sequence of nodes $h=v_0 v_1 \ldots v_k$, we define the {\em shift of $\pi$ by $h$\/} as the strategy profile $(\pi_i^h)_{i \in I}$ in which for all $i \in I$ and all histories ${w \in V^*}$, we have $\pi^h_i(w)=\pi_i(h \cdot w)$. We denote by $\outcome(\pi)^\sG_{h}$ (or simply $\outcome(\pi)_{h}$) the concatenation of $v_0 v_1 \ldots v_{k-1}$ with the one play in $\G$ that is the outcome of $\pi^h$ when starting in $v_k$. Thus, $\outcome(\pi)_{h}$ describes the outcome of a game that has somehow found itself with history $h$, and from that point, the players behave if the history had been $h$.
Given a profile
$(\pi_i)_{i\in\agentsset}$, an index ${j\in\agentsset}$, and a strategy $\pi'_j$ for Player~$j$, we use $(\pi_{-j},\pi'_j)$ to refer to the profile of strategies in which the strategy for all players but $j$ is as in $\pi$, and the strategy for Player~$j$ is $\pi'_j$. Thus,
 $(\pi_{-j},\pi'_j) = \zug{\pi_1,\pi_2,\ldots,\pi_{j-1},\pi'_j,\pi_{j+1},\ldots, \pi_n}$.



\comment{
We model reactive systems by deterministic transducers. A \emph{transducer} is a tuple $T=\la \Sigma_e, \Sigma_s, S, s_0,
\eta, L\ra$, where $\Sigma_e$ is an alphabet corresponding to the inputs generated by the environments, $\Sigma_s$ is an alphabet corresponding to the system's outputs, $S$ is a set of states, $s_0$ is an initial state, $\eta:S \times \Sigma_e\rightarrow S$ is a transition function, and $L:S\rightarrow \Sigma_s$ is a labeling function. When the system is in state $s\in S$ and it
reads a letter $\sigma_e\in\Sigma_e$ it changes its state to $s'=\delta(s,\sigma_e)$ and outputs the letter $L(s')$. Note that a transducer induces a strategy $\pi_T:(\Sigma_e)^* \rightarrow \Sigma_s$.}


\mSection{Rational Synthesis}\label{sec:rationalsynt}
In this section we define the problem of rational synthesis.
We work with the following model:
the world consists of  the {\em system} and a set of $n$ agents
$\player{1},\ldots,\player{n}$.
For uniformity we refer to the system as $\player{0}$.
We assume that Agent~$i$ controls a set $X_i$ of
variables, and the different sets are pairwise disjoint.
At each point in time, each agent sets his variables to certain
values. Thus, an action of $\player{i}$ amounts to assigning values to his variables. Accordingly, the set of actions of  \player{i} is given by $2^{X_i}$.
We use $X$ to denote $\bigcup_{0\leq i\leq n} X_i$. We use $X_{-i}$ to denote $X \setminus X_i$ for $0\leq i\leq n$.
Each of the agents (including the system) has an objective.
The objective of an agent is formulated using a linear temporal logic
formula (\ltl~\cite{Pnu77}) over the set of variables of all agents.\footnote{We could have worked with any other $\omega$-regular formalism for specifying the objectives. We chose LTL for simplicity of the presentation.}
We use $\varphi_i$ to denote the objective of \player{i}.

This setting induces the game arena $\G= \la V,v_0, \agentsset,(\Actions_i)_{i\in\agentsset}, (\Gamma_i)_{i\in\agentsset},\delta \ra$ defined as follows. The set of players $I=\{0,1,\ldots,n\}$ consists of the system and the agents. The moves of agent $i$ are all the possible assignments to its variables. Thus, $\Sigma_i = 2^{X_i}$.
We use $\Sigma$, $\Sigma_i$, and $\Sigma_{-i}$ to denote the sets $2^{X}$, $2^{X_i}$,  and $2^{X_{-i}}$, respectively.
An agent can set his variables as he wishes throughout the game. Thus $\Gamma_i(v)=\Sigma_i$ for every $v\in V$. The game records in its vertices all the actions taken by the agents so far. Hence, $V= \Sigma^*$ and for all ${v \in \Sigma^*}$ and ${\zug{\sigma_0,\ldots,\sigma_n} \in \Sigma}$, we have ${\delta(v,\sigma_0,\ldots,\sigma_n)=v \cdot\la\sigma_0, \ldots, \sigma_n\ra}$.


At each moment in time, the system gets as input an assignment in $\Sigma_{-0}$ and it generates as output an assignment in $\Sigma_0$.
For every possible history $h \in (\Sigma_{-0} \cup \Sigma_0)^*$ the system should decide what $\sigma_0\in\Sigma_0$ it outputs next.
Thus, a strategy for the system is  a function ${\pi_0:\Sigma^* \rightarrow \Sigma_0}$ (recall that $\Sigma = \Sigma_{-0}\cup\Sigma_0$ and note that indeed $V^+=\Sigma^*$). In the standard synthesis problem, we say that $\pi_0$ realizes $\varphi_0$ if all the computations that $\pi_0$ generates satisfy $\varphi_0$. In rational synthesis, on the other hand, we also generate strategies for the other agents, and the single computation that is the outcome of all the strategies should satisfy $\varphi_0$.
That is, we require $\outcome(\pi)^{\G} \models \f_0$ where $\G$ is as defined above.
In addition, we should generate the strategies for the other agents in a way that would guarantee that they indeed adhere to their strategies.

Recall that while we control the system, we have no control on the behaviors of $\player{1},\ldots,\player{n}$. Let ${\pi_0:\Sigma^* \rightarrow \Sigma_0}$ be a  strategy for the system in $\G$. Then, $\pi_0$ induces the game
${\G_{\pi_0}=\zug{\Sigma^*,\epsilon,\agentsset,(\Sigma_i)_{i\in\agentsset},
(\Gamma'_i)_{i\in\agentsset},\delta}}$, where for $i\in \agentsset \setminus\{0\}$, we have $\Gamma'_i=\Gamma_i$, and $\Gamma'_0(w)=\{\pi_0(w_{-0})\}$, where $w_{-0}$ is obtained form $w$ by projecting its letters on $\Sigma_{-0}$. Recall that $\delta$ is restricted to the relevant domain. Thus, as $\Gamma'_0$ is deterministic, we can regard $\G_{\pi_0}$ as an $n$-player (rather than $n+1$-player) game. Note that
 $\G_{\pi_0}$ contains all the possible behaviors of $\player{1},\ldots,\player{n}$, when the system adheres to $\pi_0$.

\comment{
Let $\agentssetwo=\{1,\ldots,n\}$.
Then, $T$ restricts $\G$ to the game arena $\G_T=\la \Sigma_{-0} \times S,(\epsilon,s_0),\agentssetwo,(\Sigma_i)_{i\in\agentssetwo},(\Gamma_i)_{i\in\agentssetwo},\delta_T \ra$ where $\delta_T$ is defined as follows.
Let $s$ be a state in  $S$, $w$ a word over $\Sigma_{-0}$,  $\sigma=\la \sigma_1,\ldots,\sigma_n\ra$ and $\sigma'=\la\sigma'_1,\ldots,\sigma'_n\ra$ letters over $\Sigma_{-0}$. Then
$\delta_T(({w}\sigma, s),\sigma')=({w\sigma\sigma'}, \eta(s,\sigma))$.
Thus, $\G_T$ contains all the possible behaviors of $\player{1},\ldots,\player{n}$, when the system behaves as specified in $T$.
}

\begin{definition}{Rational Synthesis}\label{def:rationalsynt}
Consider a solution concept $\gamma$.
The problem of rational synthesis (with solution concept $\gamma$) is to return,
given LTL formulas $\fsystem,\f_1,\ldots,\f_n$, specifying
the objectives of the system and the agents constituting its
environment,  a strategy profile $\pi=\la\pi_0,\pi_1,\ldots,\pi_n\ra
\in \Pi_0 \times \Pi_1 \times \cdots \times \Pi_n$  such that both (a)
$\outcome(\pi)^{\G} \models \fsystem$ and (b)
the strategy profile $\zug{\pi_1,\ldots,\pi_n}$ is a solution in the
game $\G_{\pi_0}$ with respect to the solution concept $\gamma$.
\end{definition}

The rational-synthesis problem gets a solution concept as a parameter. As discussed in Section~\ref{intro}, the fact $\zug{\pi_1,\ldots,\pi_n}$ is a solution with respect to the concept guarantees that it is not worthwhile for the agents constituting the environment to deviate from the strategies assigned to them. Several solution concepts are studied and motivated in game theory. We focus on three leading concepts, and
we first recall their
definitions and motivations in game theory.
The common setting in game theory is that the objective for each player is
to maximize his {\em payoff} -- a real
number that is a function of the play. We use $\payoff_i:\Plays\to \mathbb{R}$ to denote the payoff function of player $i$. That is, $\payoff_i$ assigns to each possible play $p$ a real number $\payoff_i(p)$ expressing the payoff of $i$ on $p$. For a strategy profile $\pi$ we use (with a slight abuse of notation) $\payoff_i(\pi)$ to abbreviate $\payoff_i(\outcome(\pi))$.

The simplest and most appealing solution concept is dominant-strategies solution.
A {\em dominant strategy} is a strategy that a player can never lose
by adhering to, regardless of the strategies of the other players.
Therefore, if there is a profile of strategies $\pi$ in which
all strategies $\pi_i$ are dominant, then no player has an
incentive to deviate from the strategy assigned to him in $\pi$.
Formally, $\pi$ is a \emph{dominant strategy profile} if
for every $1 \leq i \leq n$ and for every profile $\pi'$ with
${\pi'_i \neq \pi_i}$, we have that $\payoff_i(\pi') \leq \payoff_i(\pi'_{-i},\pi_i)$.
%
%
Consider, for example, a game played by three players: Alice, Bob and Charlie whose actions are $\{\actiona,\actionA\}$, $\{\actionb,\actionB\}$ and $\{\actionc,\actionC\}$, respectively.
 The game is played on the game arena depicted in the left of
 Figure~\ref{fig:game}. The labels on the edges are marked by the
 possible action moves. Each player wants to visit infinitely often a
 node marked by his initial letter.
In this game, Bob's strategy of choosing $\actionb$ from Node $2$ is a dominant strategy.  All of the strategies of Charlie are dominating. Alice, though, has no dominating strategy. 
Unfortunately, in many games some agents do not have
dominant strategies, thus no dominant-strategy solution exists.
Naturally, if no dominant strategy solution exists, one would still
like to consider other solution concepts.

Another well known solution concept is Nash equilibrium~\cite{Nas50}. 
A strategy profile is  {\em Nash equilibrium} if no player
has an incentive
to deviate from his strategy in $\pi$ provided he assumes the other
players adhere to the strategies assigned to them in $\pi$.
Formally, $\pi$ is  a {\em Nash equilibrium profile} if
for every $1 \leq i \leq n$ and for every  strategy
${\pi'_i \neq  \pi_i}$, we have that $\payoff_i(\pi_{-i},\pi'_i) \leq \payoff_i(\pi)$.
%
For example, the strategy profile depicted in the middle of Figure~\ref{fig:game} by dotted edges is a Nash equilibrium of the game to its left. Knowing the strategy of the other players, each player cannot gain by deviating from his strategy.

An important advantage of Nash equilibrium is that a Nash equilibrium
exists in almost every game \cite{OR94}.\footnote{In particular, all $n$-player turn-based games with $\omega$-regular objectives have Nash equilibrium ~\cite{CMJ04}.}
A weakness of Nash equilibrium is that it is not nearly as stable as
a dominant-strategy solution: if one of the other players
deviates from his assigned strategy, nothing is guaranteed.

Nash equilibrium is suited to a type of games in which the players make
all their decisions without knowledge of other players
choices.
The type of games considered in rational synthesis, however, are different, as players do have knowledge about the choices of the other players in earlier rounds of the game. 
%
To see  the problem that this setting poses for Nash
equilibrium, let us consider the {\sc Ultimatum} game.
In {\sc Ultimatum}, Player~1 chooses a value $x\in
[0,1]$, and then Player~2 chooses whether to accept the
choice, in which case the payoff of Player~1 is $x$ and the payoff of Player~2  is
$1-x$, or to reject the choice, in which case the payoff of both players is $0$.
One Nash equilibrium in {\sc Ultimatum} is $\pi=\zug{\pi_1,\pi_2}$ in which $\pi_1$ advises Player~1 to always choose $x=1$ and $\pi_2$ advises Player~2 to always reject.
It is not hard to see that $\pi$ is indeed a Nash equilibrium. In particular, if Player~2 assumes that Player~1 follows $\pi_1$, he has no incentive to deviate from $\pi_2$.
Still, the equilibrium is unstable. The reason is that $\pi_2$ is inherently not credible.
If Player~1 chooses $x$ smaller than $1$, 
 it is irrational for Player~2 to reject, and Player~1 has no reason to assume that Player~2 adheres to $\pi_2$.
This instability of a Nash equilibrium is especially true in a setting in which the players have information about the choices made by the other players. In particular, in {\sc Ultimatum}, Player~1 knows that Player~2 would make his choice after knowing what $x$ is.

To see  this problem in the setting of infinite games,
consider the strategy profile depicted in the right of
Figure~\ref{fig:game} by dashed edges. This profile is also a Nash
equilibrium of the game in the left of the figure.
It is, however, not very rational. The reason is
that if Alice deviates from her strategy by choosing $\actionA$ rather
than $\actiona$ then it is irrational for Bob to stick to his
strategy. Indeed, if he sticks to his strategy he does not meet his
objective, yet if he deviates and chooses $\actionb$ he does meet his
objective.

%
%
%
This instability of Nash equilibrium has been addressed in the definition of
subgame-perfect equilibrium~\cite{Sel75}. 
A strategy profile $\pi$ is in {\em subgame-perfect equilibrium (SPE)} if
for every possible history of the game, no player
has an incentive
to deviate from his strategy in $\pi$ provided he assumes the other
players adhere to the strategies assigned to them in $\pi$.
Formally, $\pi$ is an SPE
profile if for every possible history $h$ of the game, player
$1 \leq i \leq n$, and strategy ${\pi'_i \neq \pi_i}$,
we have that $\payoff_i(\pi_{-i},\pi'_i)_h \leq \payoff_i(\pi)_h$.
The dotted strategy depicted in the middle of Figure~\ref{fig:game} is a subgame-perfect equilibrium. Indeed, it is a Nash equilibrium  from every possible node of the arena, including non-reachable ones.

In the context of on-going behaviors, real-valued payoffs are a big
challenge and most works on reactive systems use Boolean
temporal-logic as a specification language.
Below we adjust the definition of the three solution concepts to the case the objectives are LTL formulas.\footnote{In Section~\ref{sec:MultiSol}, we make a step towards generalizing the framework to the multi-valued setting and consider the case the payoffs are taken from a finite distributive lattice.}
Essentially, the adjustment is done by assuming the following simple payoffs:
If the objective $\f_i$ of Agent~$i$ holds, then his payoff is $1$; otherwise his payoff is $0$. The induced solution concepts are then as followed. Consider a strategy profile $\pi=\zug{\pi_1,\ldots,\pi_n}$.
\begin{itemize}
\item
We say that $\pi$ is a \emph{dominant strategy profile} if
for every $1 \leq i \leq n$ and profile $\pi'$ with
${\pi'_i \neq \pi_i}$, if
$\outcome(\pi') \models \varphi_i$, then
$\outcome(\pi'_{-i},\pi_i) \models \varphi_i$.
\item
We say that $\pi$ is a \emph{Nash equilibrium profile} if
for every $1 \leq i \leq n$ and profile $\pi'$ with
${\pi'_i \neq \pi_i}$, if
$\outcome(\pi_{-i},\pi'_i) \models \varphi_i$, then
$\outcome(\pi) \models \varphi_i$.
\item
We say that $\pi$ is a \emph{subgame-perfect equilibrium profile} if
for every history ${h \in \Sigma^*}$, ${1 \leq i \leq n}$, and profile $\pi'$ with
${\pi'_i \neq \pi_i}$, if
${\outcome(\pi_{-i},\pi'_i)_h \models \varphi_i}$, then
${\outcome(\pi)_h \models \varphi_i}$.
\end{itemize}

\stam{
\begin{remark}
The discussed solution concepts are presented in their {\em weak\/} version. The strong version is the same only that $<$ is used instead of $\leq$ when comparing the payoffs. In the Boolean setting, we interpreted $a \leq b$ as $a \rightarrow b$. Accordingly, we interpret $a < b$ as $\neg a \wedge b$.
\comment{
Yak yak also weak and strong
There may be more solution concept that are reasonable in the Boolean setting. For example, requiring in addition to one of the solutions above that all objectives are met when all agents adhere to their strategy. Formally, requiring that  $\outcome(\pi) \models \bigwedge_{i\in[0..n]} \varphi_i$. That might indeed be a too strong requirement in certain cases. In the context of one agent and a system another reasonable solution to consider, in view of the system's distinguished role, is to interpret $\prec$ as ``if  the objective of the system does not hold because the agent deviated from his strategy, then the agent's objective no longer holds as well". Formally, the profile $(\pi_0,\pi_1)$ will adhere to this if $\outcome(\pi_0,\pi_1) \models \f_0 \wedge \f_1$ and
$\forall \pi'_1. {\outcome(\pi_0,\pi'_1) \models \neg \fsystem
\implies
\outcome(\pi_0,\pi'_1) \models \neg \f_i}$.
}
\end{remark}
}

\mSection{Solution in the Boolean Setting}\label{sec:BoolSol}

In this section we solve the rational-synthesis problem.
Let $\agentsset = \{0,1,\ldots,n\}$ denote the set of agents.
Recall that $\Actions_i=2^{X_i}$ and $\Actions=2^X$, where $X= \cup_{i\in\agentsset}X_i$, and that the partition of the variables among the agents induces a game arena with states in $\Sigma^*$.
%
Expressing rational synthesis involves properties of strategies and histories.
\emph{Strategy Logic}~\cite{CHP07} is a logic that treats strategies in games as explicit first-order objects. Given an LTL formula $\psi$ and strategy variables $z_0,\ldots,z_n$ ranging over strategies of the agents, the strategy logic formula $\psi(z_0,\ldots,z_n)$ states that $\psi$ holds in the outcome of the game in which Agent $i$ adheres to the strategy $z_i$. The use of existential and universal quantifiers on strategy variables enables strategy logic to state that a given profile consists of dominant strategies or is a Nash equilibrium. However, strategy logic is not strong enough to state the existence of a subgame perfect equilibrium. The reason is that a formula ${\f(z_0,\ldots,z_n)}$ in strategy logic assumes that the strategies $z_0,\ldots,$ $z_n$ are computed from the initial vertex of the game, and it cannot refer to histories that diverge from the strategies. We therefore extend strategy logic with first order variables that range over arbitrary histories of the game.

\mSubSection{Extended Strategy Logic}
Formulas of
\emph{Extended Strategy Logic} (\ESL) are defined with respect to a game   $\G= \la V,v_0, \agentsset,(\Actions_i)_{i\in\agentsset}, (\Gamma_i)_{i\in\agentsset},$ $\delta \ra$, a set $\HistoryVars$ of history variables,
 and sets $\StrategyVars_i$ of strategy variables for $i\in\agentsset$.
 Let $I =\{0,\ldots,n\}$,
 $\Sigma = \Sigma_0 \times \cdots \times \Sigma_n$, and
let $\psi$ be an LTL formula over $\Sigma$. Let $\h$ be a history variable in $\HistoryVars$, and let
$\z_0,...,\z_n$ be strategy variables in $\StrategyVars_0,\ldots,\StrategyVars_n$, respectively. We use $z$ as an abbreviation for $\z_0,...,\z_n$.
The set of \ESL\ formulas is defined inductively as
follows.\footnote{We note that strategy logic as defined
  in~\cite{CHP07} allows the application of LTL path operators
  ($\Next$ and $\Until$) on strategy logic closed formulas.
  Since we could not come up with a meaningful specification  that
  uses such applications, we chose to ease the presentation and do
  not allow them in ESL. Technically, it is easy to extend ESL and
  allow such applications.}
$$\begin{array}{ll}\Psi ::= & \psi(z) ~|~
\psi(z;h) ~|~ \Psi \vee \Psi ~|~   \neg \Psi ~|~ \exists z_i.\Psi ~|~ \eexists h.\Psi \end{array}$$
We use the usual abbreviations $\wedge, \rightarrow$, and $\forall$. We denote by $\Free(\Psi)$ the set of strategy and history variables that are {\em free} (not in a scope of a quantifier) in $\Psi$. A formula $\Psi$ is  \emph{closed} if $\Free(\Psi)=\emptyset$. The {\em alternation depth\/} of a variable of a closed formula is the number of quantifier switches ($\exists\forall$ or $\forall\exists$, in case the formula is in positive normal form) that bind the variable. The {\em alternation depth\/} of closed formula $\Psi$ is the maximum alternation depth of a variable occurring in the formula.

We now define the semantics of ESL.
Intuitively, an ESL formula of the form $\psi(z;h)$ is interpreted over the game whose prefix matches the  history $h$ and the suffix starting where $h$ ends is the outcome of the game that starts at the last vertex of $h$ and along which each agent ${i\in\agentsset}$ adheres to his strategy in $z$.
Let $\X \subseteq \HistoryVars  \cup \bigcup_{i\in\agentsset} \StrategyVars_i$ be a set of variables. An assignment $\A_\X$ assigns to every history variable $h\in\X \cap \HistoryVars$, a history \comment{$\A_\X(h) \in \Sigma^*$}$\A_\X(h) \in V^+$ and
assigns to every strategy variable $\z_i \in \X \cap \StrategyVars_i$, a strategy $\A_\X(\z_i) \in\Pi_i$.
 Given an assignment $\A_\X$ and a strategy $\pi_i \in \Pi_i$, we denote by $\A_\X[\z_i\!\leftarrow\!\pi_i]$ the assignment $\A'_{\X \cup \{\z_i\}}$ in which ${\A'_{\X\cup\{ \z_i \}}(z_i) = \pi_i}$ and for a variable ${x \neq z_i}$ we have ${\A'_{\X\cup\{ \z_i \}}(x) = \A_\X(x)}$.  For histories of the game ${w\in V^+}$\comment{$w\in \Sigma^*$} we define  $\A_\X[h\! \leftarrow\!w]$  similarly.

We now describe when a given game $\G$ and a given assignment $\A_\X$ satisfy an \ESL\ formula $\Psi$, where $\X$ is such that $\Free(\Psi) \subseteq \X$. For LTL, the semantics is as usual~\cite{MP92}.
\vspace{-4mm}
\bc
\begin{tabular}{l@{\ \ }c@{\ \ \ }l@{\quad\quad\quad}l@{\ \ }c@{\ \ \ }l}
       \hspace{-3mm} $(\G,\!\A_\X) \models 	\psi(\z)$ & iff &
       $\outcome(\A_\X(\z))^\sG \models \psi$	
       &
       \hspace{-3mm} $(\G,\!\A_\X) \models \Psi_1 \vee \Psi_2$ & iff &
        $(\G,\!\A_\X ) \models \Psi_1$ or
        $(\G,\!\A_\X ) \models \Psi_2$
    \\[2mm]
        \hspace{-3mm} $(\G,\!\A_\X) \models 	\psi(\z;h)$ & iff &
        $\outcome(\A_\X(\z))^\sG_{\A_\X(h)} \models \psi$	
        &
        \hspace{-3mm} $( \G,\!\A_\X ) \models \exists \z_i. \Psi$ & iff &
        $\exists \pi_i\tightin\Pi_i.(\G,\!\A_\X[\z_i\!\leftarrow\!\pi_i]) \models \Psi$
    \\[1mm]
      \hspace{-3mm} $( \G,\!\A_\X ) \models \neg\Psi$ & iff &
        $(\G,\!\A_\X) \nmodels \Psi$
        &
        \hspace{-3mm} $( \G,\!\A_\X ) \models \eexists h. \Psi$ & iff &
        $\exists w\tightin V^+.(\G,\!\A_\X[h\!\leftarrow\!w]) \models \Psi$
    \\
\end{tabular}\\[2mm]
\ec

\comment{
\bc
\begin{tabular}{l@{\quad}c@{\quad}l}
       \hspace{-3mm} $(\G,\!\A_\X) \models 	\psi(\z)$ & iff &
       $\outcome(\A_\X(\z))^\sG \models \psi$	
    \\[1mm]
        \hspace{-3mm} $(\G,\!\A_\X) \models 	\psi(\z;h)$ & iff &
        $\outcome(\A_\X(\z))^\sG_{\A_\X(h)} \models \psi$	
    \\[1mm]
        \hspace{-3mm} $(\G,\!\A_\X) \models \Psi_1 \vee \Psi_2$ & iff &
        $(\G,\!\A_\X ) \models \Psi_1$ or
        $(\G,\!\A_\X ) \models \Psi_2$
    \\
      \hspace{-3mm} $( \G,\!\A_\X ) \models \neg\Psi$ & iff &
        $(\G,\!\A_\X) \nmodels \Psi$
    \\[1mm]
          \hspace{-3mm} $( \G,\!\A_\X ) \models \exists \z_i. \Psi$ & iff &
        $\exists \pi_i\tightin\Pi_i.(\G,\!\A_\X[\z_i\!\leftarrow\!\pi_i]) \models \Psi$
    \\[1mm]
        \hspace{-3mm} $( \G,\!\A_\X ) \models \eexists h. \Psi$ & iff &
        $\exists w\tightin V^+.(\G,\!\A_\X[h\!\leftarrow\!w]) \models \Psi$
    \\
\end{tabular}\\[2mm]
\ec}

For  an \ESL\  formula $\Psi$  we use $\sema{\Psi}$ to denote its set of satisfying assignments; that is, $\sema{\Psi} = \{ (\G,\A_\X)   ~|~  \X = \Free(\Psi) \mbox{ and } ( \G,\!\A_\X ) \models \Psi\}$.
Given an \ESL\ formula $\Psi$ and a game graph $\G$, we denote by $\sema{\Psi}_\G$ the assignment $\A_\X$ to the free variables in $\Psi$ such that $(\G,\A_\X) \in \sema{\Psi}$.


Before we show how $\sema{\Psi}_\G$ can be computed 
 we
show that \ESL\ is strong enough to express the solution to the rational-synthesis problems for the three solution concepts we study.

\mSubSection{Expressing Rational Synthesis}\label{sec:expressolutions}
We now show that the rational synthesis problem for the three solution concepts we study can be stated in \ESL.
We first state that a given strategy profile $y = (y_i)_{i\in I}$ is a solution concept on the game $\G_{y_0}$, that is, the game induced by $\G$ when Agent $0$ adheres to his strategy in $y$.
We use $\agentssetwo$ to denote the set  $\{1,\ldots,n\}$, that is, the set of all agents except for the system, which is Agent $0$. Given a strategy profile $z=(z_i)_{i\in\agentsset}$, we use $(z_{-\{i,0\}},y_i,y_0)$ to denote the strategy profile where all agents but $i$ and $0$ follow $z$ and agents $i$ and $0$ follow $y_i$ and $y_0$, respectively.
For $i\in\agentsset$, let $\f_i$ be the objective of Agent~$i$.
For a solution concept $\gamma\in\{\dominatingge,\nashge,\spege\}$ and a strategy profile $y=(y_i)_{i\in\agentsset}$, the formula $\Psi^\gamma(y)$, expressing that the profile $(y_i)_{i\in\agentssetwo}$ is a solution with respect to $\gamma$ in $\G_{y_0}$, is defined as follows.\\[2mm]
 $\bullet$ $\Psi^{\scriptsize{\dominatingge}}(y) := \bigwedge_{i\in\agentssetwo} \forall z.\ (\f_i(z_{-0},y_0) \rightarrow \f_i(z_{-\{i,0\}},y_i,y_0)).$
\\[1mm]
 $\bullet$ $\Psi^{\scriptsize{\nashge}}(y) := \bigwedge_{i\in\agentssetwo} \forall \z_i.\ (\f_i(y_{-i},\z_i)\!\rightarrow\!\f_i(y)).$
\\[1mm]
 $\bullet$ $\Psi^{\scriptsize{\spege}}(y) := \fforall \h. \bigwedge_{i\in\agentssetwo} \forall \z_i.\ ((\f_i(y_{-i},\z_i,\h)\!\rightarrow\!(\f_i(y,\h)).$ \\

We can now state the existence of a solution to the rational-synthesis problem with input $\f_0,\ldots,\f_n$ by the closed formula $\Phi^\gamma := \exists (y_i)_{i\in\agentsset}.(\f_0 ((y_i)_{i\in\agentsset})\wedge \Psi^\gamma((y_i)_{i\in\agentsset}))$. Indeed, the formula specifies the existence of a strategy profile whose outcome satisfies $\f_0$ and for which the strategies for the agents in $\agentssetwo$ constitute a solution with respect to $\gamma$ in the game induced by $y_0$.

\mSubSection{ESL Decidability}

In order to solve the rational-synthesis problem we are going to use automata on infinite trees.
Given a set $D$ of directions, a {\em $D$-tree\/} is the set $D^*$. The elements in $D^*$ are the {\em nodes} of the tree. The node $\epsilon$ is the root of the tree. For a node $u \in D^*$ and a direction $d \in D$, the node $u \cdot d$ is the {\em successor} of $u$ with {\em direction\/} $d$. Given $D$ and an alphabet $\Sigma$,
a $\Sigma$-labeled $D$-tree is a pair $\la D^*,\tau \ra$ such that
$\tau:D^* \rightarrow \Sigma$ maps each node of $D^*$ to a letter in $\Sigma$.

An \emph{alternating parity tree automaton (\APT)}  is a tuple
$\A=\zug{\Sigma,D,Q,\delta_{0},\delta,\chi}$,
where $\Sigma$ is the input
alphabet, $D$ is the directions set, $Q$ is a finite set of states,
$\delta_0$ is the initial condition, $\delta$ is the transition relation and $\chi: Q \mapsto \{1,\ldots,k\}$ is the parity condition. The initial condition $\delta_0$ is a positive boolean formula over $Q$ specifying the initial condition. For example, $(q_1 \vee q_2) \wedge q_3$ specifies that the \APT\ accepts the input tree if it accepts it from state $q_3$ as well as from $q_1$ or $q_2$. The transition function $\delta$  maps each state and letter to a boolean formula over $D\times Q$. Thus, as with $\delta_0$, the idea is to allow the automaton to send copies of itself in different states. In $\delta$, the copies are sent to the successors of the current node, thus each state is paired with the direction to which the copy should proceed. Due to the lack of space, we refer the reader to~\cite{GTW02} for the definition of runs and acceptance. 

Base ESL formulas, of the form $\psi(z,h)$, refer to exactly one
strategy variable for each agent, and one history variable. The
assignment for these variables can be described by a $(\Sigma \times
\{\bot,\top\})$-labeled $\Sigma$-tree, where
the $\Sigma$-component of the labels is used in order to describe the strategy profile $\pi$ assigned to the strategy variable, and the $\{\bot,\top\}$-component of the labels is used in order to label the tree by a unique finite path corresponding to the history variable. We refer to a $(\Sigma \times \{\bot,\top\})$-labeled $\Sigma$-tree as a \emph{strategy-history tree}.
A node $u=d_0 d_1 \ldots d_k$ in a strategy-history tree $\zug{\Sigma^*,\tau}$ corresponds to a history of the play in which at time $0\leq j \leq k$, the agents played as recorded in $d_j$. A label $\tau(u)=(\sigma_0,\ldots,\sigma_n,\dashv)$ of node $u$ describes (1) for each agent $i$, an action $\sigma_i$ that the strategy $\pi_i$ advises Agent~$i$ to take when the history of the game so far is $u$, and (2) whether the node is along the path corresponding to the history. Among the $|\Sigma|$ successors of $u$ in the strategy-history tree, only the successor $u \cdot \tau(u)$ corresponds to a scenario in which all the agents adhere to  their strategies in the strategy profile described in $\zug{\Sigma^*,\tau}$.
We say that a path $\rho$ in $\zug{\Sigma^*,\tau}$ is \emph{obedient} if for all nodes $u \cdot d \in \rho$, for $u \in \Sigma^*$ and $d \in \Sigma$, we have $d=\tau(u)$. Note that  there is a single obedient path in every strategy tree. This path corresponds to the single play in which all agents adhere to their strategies. The $\{\bot,\top\}$ labeling is legal if there is a unique finite path starting at the root, all of whose node are marked with $\top$. Note that there is a single path in the tree whose prefix is marked by $\top$'s and whose suffix is obedient.

An ESL formula $\Psi$ may contain several base formulas. Therefore, $\Psi$ may contain, for each $i \in I$, several strategy variables in
 $\StrategyVars_i$ and several history variables in $\HistoryVars$. For $i \in I$, let $\{z_i^1,\ldots,z_i^{m_i}\}$ be the set of strategy variables in $\Psi \cap \StrategyVars_i$. Recall that each strategy variable $z_i^j \in \StrategyVars_i$ corresponds to a strategy $\pi_i^j:\Sigma^* \rightarrow \Sigma_i$.
Let $\{h_1,\ldots,h_m\}$ be the set of history variables in $\Psi$.
Recall that each history variable $h$ corresponds to a word in $\Sigma^*$, which can be seen as a function $w_h:\Sigma^* \rightarrow \{\top,\bot\}$ labeling only that word with $\top$'s.
 Thus, we can describe an assignment to all the variables in $\Psi$ by a $\Tau$-labeled $\Sigma$-tree, with $\Tau=\Sigma_0^{m_0} \times \Sigma_1^{m_1} \times \cdots \times \Sigma_n^{m_n} \times \{\bot,\top\}^m$.

We  solve the rational synthesis problem
using tree automata that run on $\Tau$-labeled $\Sigma$-trees. Note that the specification of rational synthesis involves an external quantification of a strategy profile.  We construct an automaton $\U$ that accepts all trees that describe a strategy profile that  meets the desired solution. A witness to the nonemptiness of the automaton then induces the desired strategies.
%

%

We define $\U$ as an \APT.
Consider an ESL formula $\psi(\z,\h)$. Consider a strategy tree $\zug{\Sigma^*,\tau}$. Recall that $\psi$ should hold along the path that starts at the root of the tree, goes through $\h$, and then continues to $\outcome(z)_h$. Thus, adding history variables to strategy logic results in a {\em memoryful logic} \cite{KV06a}, in which LTL formulas have to be evaluated not along a path that starts at the present, but along a path that starts at the root and goes through the present. The memoryful semantics imposes a real challenge on the decidability problem, as one has to follow all the possible runs of a nondeterministic automaton for $\psi$, which involves a satellite implementing the subset construction of this automaton \cite{KV06a}. Here, we use instead the $\{\bot,\top\}$-component of the label of $\tau$.

The definition of the \APT $\A_\Psi$ for $\sema{\Psi}_\G$ works by induction on the structure of $\Psi$.
At the base level, we have formulas of the form $\psi(\z,\h)$, where $\psi$ is an LTL formula, $\z$ is a strategy profile, and $\h$ is a history variable. The constructed automaton then has three tasks.
The first task is to check that the $\{\bot,\top\}$ labeling is legal; i.e. there is a unique path in the tree marked by $\top$'s. The second task is to detect the single path that goes through $\h$ and continues from $\h$ according to the strategy profile $z$. The third task is to check that this path satisfies $\psi$.
The inductive steps then built on \APT\ complementation, intersection, union and projection~\cite{MS87}. In particular, as in strategy logic, quantification over a strategy variable for agent $i$ is done by ``projecting out" the corresponding $\Sigma_i$ label from the tree. That is, given an automaton $\A$ for $\Psi$, the automaton for $\exists z_i. \Psi$ ignores the $\Sigma_i$ component that refers to $z_i$ and checks $\A$ on a tree where this component is guessed.
The quantification over history variables is similar. Given an automaton $\A$ for $\Psi$ the automaton for $\eexists h. \Psi$ ignores the $\{\bot,\top\}$ part of the label that corresponds to $h$ and checks $\A$ on a tree where the $\{\bot,\top\}$ part of the label is guessed.


\BT \label{thm:eslcompl}
Let $\Psi$ be an \ESL\ formula over $\G$. Let $d$ be the alternation depth of $\Psi$. 
 We can construct an \APT\ $\A_\Psi$ such that $\A_\Psi$ accepts $\sema{\Psi}_\G$ and its emptiness can be checked in time $(d+1)$-EXPTIME in the size of $\Psi$.
\ET


\mSubSection{Solving Rational Synthesis}
We can now reduce rational-synthesis to \APT\ emptiness.

\BT\label{thm:nashge} The LTL rational-synthesis problem is 2EXPTIME-complete for the solution concepts of dominant strategy, Nash equilibrium, and subgame-perfect equilibrium.
\ET

\BPF
We have shown in Section~\ref{sec:expressolutions} that the rational-synthesis problem for $\gamma \in \{\dominatingge,\nashge,\spege\}$ can be specified by an \ESL\ formula $\Phi^\gamma$ with one alternation. It follows from Theorem~\ref{thm:eslcompl} that we can construct an \APT\ accepting $\sema{\Phi^\gamma}_\G$ (where $\G$ is as defined in Section~\ref{sec:rationalsynt}) whose emptiness can be solved in 2EXPTIME.
Hence, the problem is in 2EXPTIME.

Hardness in 2EXPTIME follows easily from the 2EXPTIME-hardness of LTL synthesis \cite{Ros92}. Indeed, synthesis against a hostile environment can be reduced to rational synthesis against an agent whose objective is {\em true}.
\EPF

\begin{remark}\label{rem:control} In the above we have shown how to solve the problem of rational synthesis.
It is easy to extend our algorithm to solve the problem of {\em rational control}, where one needs to control a system in a way it would satisfy its specification assuming its environment consists of rational agents whose objectives are given.
Technically, the control setting induces the game to start with, thus the strategy trees are no longer $\Sigma$-trees, and rather they are $(S \times \Sigma)$-trees, where $S$ is the state space of the system we wish to control.
\stam{
In this case we can think of the game graph as a $V$-labeled $\Sigma$-tree where $V$ are the nodes of the graph. A strategy for agent $i$ from $V^+$ to $\Sigma_i$ can be enhanced to a function from $(V\times\Sigma)^+$ to $\Sigma_i$, and similarly for a strategy profile. Thus, instead of building tree automata that run on $\Sigma$-labeled $\Sigma$-trees (or $\Sigma\times\{\bot,\top\}$-labeled $\Sigma$-trees) we can work with automata running over $\Sigma \times V$-labeled $\Sigma$-trees (or $\Sigma\times V \times \{\bot,\top\}$-labeled $\Sigma$-trees).
}
\end{remark}

\mSection{Solution in the  Multi-Valued  Setting}\label{sec:MultiSol}
As discussed in Section~\ref{intro}, classical applications of game
theory consider games with quantitative payoffs. The extension of
the synthesis problem to the rational setting calls also for an
extension to the quantitative setting.
Unfortunately, the full quantitative setting
is undecidable already in the context of model checking~\cite{CCHKM05}.
In this section we study a decidable fragment of the quantitative
rational synthesis problem: the payoffs are taken from {\em finite
De-Morgan lattices}. A lattice
$\zug{A,\leq}$ is a partially ordered set in which every two elements
$a,b \in A$ have a least upper bound ($a$ \emph{join} $b$, denoted $a\join b$) and a greatest
lower bound ($a$ \emph{meet} $b$, denoted $a\meet b$).
A lattice is {\em distributive} if for every $a,b,c\in A$, we have
$a\meet (b\join c) = (a\meet b)\join (a\meet c)$. De-Morgan lattices
are distributive lattices in which every element $a$ has a unique
complement element $\lnot a$ such
that $\lnot\lnot a = a$,  De-Morgan rules hold, and $a \le b$ implies
$\lnot b \le \lnot a$. Many useful payoffs are taken from finite
De-Morgan lattices: all payoffs that are linearly ordered, payoffs
corresponding to subsets of some set, payoffs corresponding to
multiple view-points, and more \cite{KL07,KL07b}.

\stam{
The first restriction we impose is that we consider the multi-valued
setting of finite De-Morgan Lattices.
The second restriction we impose on the setting is that we assume all
the specifications and objectives involved are given in the latticed
version of deterministic \buchi\  automata (namely, LDBW).
We impose this restriction since currently there is no satisfactory
generalization of determinization for LNBW, and the technical issues
involved are orthogonal to these that are challenged by rational synthesis.
Finally, we consider here only Nash equilibrium.

We assume familiarity with the definitions and
notation of latticed automata and $\omega$-games as presented in
\cite{KL07,KL07b}.
}
We specify qualitative specifications using the temporal logic {\em
  latticed LTL} (LLTL, for short), where the truth
value of a specification is an element in a lattice.
For a strategy profile $\pi$ and an LLTL objective $\varphi_i$ of
Agent~$i$, the payoff of Agent~$i$ in $\pi$ is the truth value of
$\varphi_i$
in $\outcome(\pi)$.
A synthesizer would like to find a profile $\pi$ in which
$\payoff_0(\pi)$ is as high as possible. Accordingly, we define the
latticed rational synthesis as follows.

\begin{definition}{Latticed Rational Synthesis}
Consider a solution concept $\gamma$.
The problem of latticed rational synthesis (with solution concept
$\gamma$) is to return,
given LLTL formulas $\varphi_0,\ldots,\varphi_n$ and a lattice value
$v\in \L$, a strategy profile $\pi=\zug{\pi_0,\pi_1,\ldots,\pi_n} \in
\Pi_0 \times \Pi_1 \times \cdots \times \Pi_n$  such that (a)
$\payoff_0(\pi) \geq v$ and (b)
the strategy profile $\zug{\pi_1,\ldots,\pi_n}$ is a solution in the
game $\G_{\pi_0}$ with respect to the solution concept $\gamma$.
\end{definition}

In the Boolean setting, we reduced the
rational-synthesis problem to
decidability of  \ESL. The decision procedure for \ESL\
is based on the automata-theoretic approach, and specifically on
APT's.
In the lattice setting, automata-theoretic machinery is not as
developed as in the Boolean case.
Consequently, we restrict
attention to LLTL specifications that can be translated to
deterministic latticed B\"uchi word automata (LDBW), and to the
solution concept of Nash equilibrium.\footnote{A {\em B\"uchi\/} acceptance
  conditions specifies a subset $F$ of the states, and an infinite
  sequence of states satisfies the condition if it visits $F$
  infinitely often. A {\em generalized B\"uchi condition\/} specifies
  several such sets, all of which should be visited infinitely often.}

An LDBW can be expanded into
a deterministic latticed B\"uchi tree automata (LDBT), which is the
key behind the analysis of strategy trees.
It is not hard to lift to the latticed setting almost all the other
operations on tree automata that are needed in order to solve rational
synthesis. An exception is the problem of emptiness.
In the Boolean case, tree-automata emptiness is
reduced to deciding a two-player game \cite{GH82}. Such games are
played between an $\vee$-player, who has a winning strategy iff the
automaton is not empty (essentially, the $\vee$-player chooses the
transitions with which the automaton accepts a witness tree), and a
$\wedge$-player, who has a winning strategy otherwise (essentially,
the $\wedge$-player chooses a path in the tree that does not satisfy
the acceptance condition).
A winning strategy for the $\vee$-player induces a
labeled tree accepted by the tree automaton.

In latticed games, deciding a game amounts to finding a lattice value
$l$ such that the $\vee$-player can force the game to computations in
which his payoff is at least $l$. The value of the game need not be achieved
by a single strategy and algorithms for analyzing latticed
games consider values that emerge as the join of values obtained by
following different strategies \cite{KL07b,SG05}.
A labeled tree, however, relates to a single strategy. Therefore,
the emptiness problem for latticed tree automata, to which the
latticed rational synthesis is reduced, cannot be reduced to
solving latticed games. Instead, one has to consider the {\em
  single-strategy\/} variant of latticed games, namely the problem of
finding values that the $\vee$-player can ensure by a single
strategy.
We address this problem below.

\begin{theorem}
\label{single strategy}
Consider a latticed B\"uchi game $G$. Given a lattice
element $l$, we can construct a Boolean generalized-B\"uchi
game $G_l$ such that the $\vee$-player can
achieve value greater or equal $l$ in $G$ using a single strategy iff
the $\vee$-player wins in $G_l$.
The size of $G_l$ is bounded by $|G|\cdot|\L|^2$ and $G_1$ has at most
$|\L|$ acceptance sets.
\end{theorem}

Using Theorem~\ref{single strategy}, we can solve the
latticed rational synthesis problem in a fashion
similar to the one we used in the
Boolean case. We represent strategy profiles by
$\Sigma$-labeled $\Sigma$-trees, and sets of profiles by tree
automata.
We construct two Boolean generalized-\buchi\  tree automata.
The first, denoted $\A_0$, for the language of all profiles $\pi$ in which
$\payoff_0(\pi) \ge v$, and the second, denoted $\A_N$, for the
language of all Nash equilibria. The
intersection of $\A_0$ and $\A_N$ then contains all the solutions to
the latticed rational synthesis problem.
Thus, solving the problem amounts to returning a witness to the
nonemptiness of the intersection, and we have the following.

\BT\label{thm:lnashge} The latticed rational-synthesis problem for
objectives in LDBW and the solution concept of Nash equilibrium is
in EXPTIME.
\ET

We note that the lower complexity with respect to the Boolean
setting (Theorem~\ref{thm:nashge}) is
only apparent, as the objectives are given in LDBWs, which are less
succinct than LLTL formulas \cite{KL07,KV05b}.

\section{Discussion}
We introduced \emph{rational synthesis} --- synthesizing a system that
functions in a rational environment. As in traditional synthesis, one
cannot control the agents that constitute the environment. Unlike
traditional synthesis, the agents have objectives, we can suggest a strategy for each
agent, and we can assume that rational agents follow
strategies they have no incentive to deviate from.

The solution of the rational synthesis problem relies on an extension
of strategy logic~\cite{CHP07}.
The modularity of our solution separates the
game-theoretic considerations and the synthesis technique.
Indeed our technique can be applied to any solution concept that can be expressed in extended strategy logic. We show that for the common solution concepts of dominant strategies equilibrium, Nash equilibrium, and subgame perfect equilibrium, rational synthesis has the same complexity as traditional synthesis
The versatility of the extended logic
enables many extensions of the setting. For example, one can associate
different solutions concepts with different sub-specifications. In
particular, it is often desirable in practice to ensure that some
properties of the system hold regardless of the rationality of the
agents. This can be done by letting the specifier specify, in addition
to $\varphi_0$, also an LTL formula $\varphi'_0$ (typically $\varphi_0
\rightarrow \varphi'_0$) that should be satisfied in the traditional
synthesis interpretations, namely in all environments.

\small

\normalsize

 \tikzstyle{graynode}=[fill=none,circle,draw=black,text=black]
  \tikzstyle{rednode}=[fill=none,circle, draw=red,text=black]
  \tikzstyle{greennode}=[fill=none,circle, draw=green,text=black]
  \tikzstyle{bluenode}=[fill=none,circle, draw=blue,text=black]

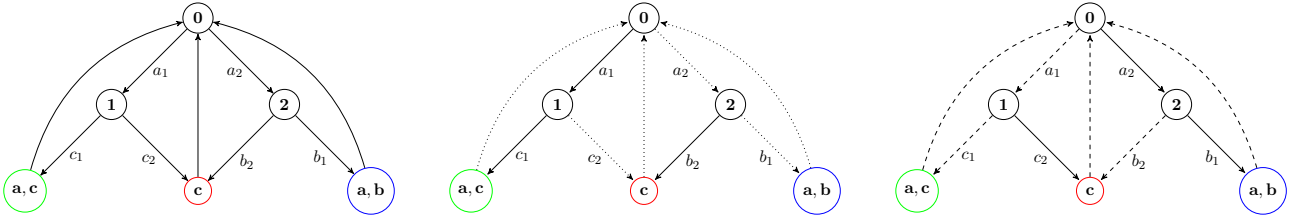
\begin{figure}
\begin{center}
\resizebox{1.0\hsize}{!}{
\begin{tikzpicture}[->,>=stealth',shorten >=.1pt,auto,node distance=2.8cm,
                    semithick]
  \node[graynode]         (one)                    {$\mathbf{1}$};
  \node[graynode]         (zero) [above right of=one] {$\mathbf{0}$};
  \node[rednode]          (four) [below right of=one] {$\mathbf{c}$};
  \node[graynode]         (two) [below right of=zero] {$\mathbf{2}$};
  \node[greennode]        (three) [below left  of=one] {${\mathbf{a},\mathbf{c}}$};
  \node[bluenode]         (five) [below right of=two] {${\mathbf{a},\mathbf{b}}$};
  \path
        (zero)  edge        [swap]      node [below] {\wactiona} (one)
                edge              node [below] {\wactionA} (two)
        (one)   edge           [swap]   node [below] {\wactionc} (three)
                edge              node [below] {\wactionC} (four)
        (two)   edge          [swap]    node [below] {\wactionB} (four)
                edge              node [below] {\wactionb} (five)
        (four)  edge   node {} (zero)
        (three) edge  [bend left] node [below] {} (zero)
        (five)  edge  [bend right] node [below] {} (zero)
        ;
\end{tikzpicture}
\quad\quad\quad
\begin{tikzpicture}[->,>=stealth',shorten >=1pt,auto,node distance=2.8cm,
                    semithick]
  \node[graynode]         (one)                    {$\mathbf{1}$};
  \node[graynode]         (zero) [above right of=one] {$\mathbf{0}$};
  \node[rednode]          (four) [below right of=one] {${\mathbf{c}}$};
  \node[graynode]         (two) [below right of=zero] {$\mathbf{2}$};
  \node[greennode]        (three) [below left  of=one] {${\mathbf{a},\mathbf{c}}$};
  \node[bluenode]         (five) [below right of=two] {${\mathbf{a},\mathbf{b}}$};
  \path
        (zero)  edge        [ swap]      node [below] {\wactiona} (one)
                edge          [dotted]    node [below] {\wactionA} (two)
        (one)   edge           [swap]   node [below] {\wactionc} (three)
                edge            [dotted]  node [below] {\wactionC} (four)
        (two)   edge          [swap]    node [below] {\wactionB} (four)
                edge            [dotted]  node [below] {\wactionb} (five)
        (four)  edge  [dotted] node {} (zero)
        (three) edge  [dotted, bend left] node {} (zero)
        (five)  edge  [dotted, bend right] node {} (zero)
        ;
\end{tikzpicture}
\quad\quad\quad
\begin{tikzpicture}[->,>=stealth',shorten >=1pt,auto,node distance=2.8cm,
                    semithick]
  \node[graynode]         (one)                    {$\mathbf{1}$};
  \node[graynode]         (zero) [above right of=one] {$\mathbf{0}$};
  \node[rednode]          (four) [below right of=one] {${\mathbf{c}}$};
  \node[graynode]         (two) [below right of=zero] {$\mathbf{2}$};
  \node[greennode]        (three) [below left  of=one] {${\mathbf{a},\mathbf{c}}$};
  \node[bluenode]         (five) [below right of=two] {${\mathbf{a},\mathbf{b}}$};
  \path
        (zero)  edge        [dashed, swap]      node [below] {\wactiona} (one)
                edge              node [below] {\wactionA} (two)
        (one)   edge           [dashed, swap]   node [below] {\wactionc} (three)
                edge              node [below] {\wactionC} (four)
        (two)   edge          [dashed, swap]    node [below] {\wactionB} (four)
                edge              node [below] {\wactionb} (five)
        (four)  edge   [dashed] node [below]  {} (zero)
        (three) edge  [dashed, bend left] node [below]  {} (zero)
        (five)  edge  [dashed, bend right] node [below] {} (zero)
        ;
\end{tikzpicture}
\quad
}
\caption{A game, two Nash equilibria and one subgame-perfect equilibrium.}\label{fig:game}
\end{center}
\end{figure}

\appendix

\section{Proofs}
\subsection{Proof of Theorem~\ref{thm:eslcompl}}

The construction proceeds by induction on the structure of $\Psi$.
Note that while the APT is defined with respect to $\Tau$-labeled $\Sigma$-trees, a base formula $\psi(z,h)$ focuses on a $(\Sigma \times \{\bot,\top\})$ projection of the label (the one assigning values to the variables in $z$ and $h$).
We describe here in detail the base case, where $\Psi = \psi(\z,\h)$. The case where $\Psi = \psi(h)$ can be derived from the case $\Psi=\psi(z,h)$ by checking in addition that only the root is labeled $\top$. The cases
$\Psi$ is of the form  $\Psi_1 \vee \Psi_2, \neg \Psi_1, \exists z_i. \Psi_1$, and $\exists h.\Psi_1$ follow from the closure of \APT s to union, complementation, and projection.


The complexity analysis follows from the fact that the automaton for $\psi(\z,\h)$ is exponential in $\psi$, and each sequence of quantifiers that increases the alternation depth by one, involves an exponential blow up in the state space and a polynomial blow up in the index \cite{MS87}. Thus,  the number of states in $\A_\Psi$ is  $(d+1)$-exponential in $\Psi$ and the index of $\A_\Psi$ is polynomial (of degree $d$) in $\Psi$,  where $d$ is the alternation depth of $\Psi$. Since the projection operation results in a nondeterministic (rather than an alternating) tree automaton, the emptiness check when the last operation is projection does not involve an additional exponential blow up.

Let $\Psi = \psi(\z,\h)$.
Given an LTL formula $\psi$, one can construct an \APT\ $\U_\psi$ with $2^{O(|\psi|)}$ states and index $3$ such that $\U_\psi$ accepts all trees all of whose paths satisfy $\psi$~\cite{VW94}.
Let $\U_\psi=\la \Actions, \Actions, Q,\delta^0, \delta, \chi\ra$. For
the first and second tasks we use four states $\ph$, $\pf$, $\ptop$,
and $\pbot$. The automaton $\A_\Psi$ starts by sending two copies, one
at the initial state of $\U_\g$ and one at $\ph$. The copy in state $\ph$ follows the \emph{history}, i.e. the path marked with $\top$ labels. When it reads a node with a $\bot$ label, marking that the history ends and the \emph{future} begins, it moves to the state $\pf$.  From the state $\pf$, this copy
checks that the agents adhere to the strategy. If a violation of the
strategy is detected, the copy concludes that $\psi$ need not be
evaluated along the path it traversed and moves to $\ptop$. If another
$\top$ has been read, the copy conclude that the
$\{\top,\bot\}$-component is illegal and moves to $\pbot$.  Formally,
$\A_\Psi = \la \Actions \times \{\bot,\top\}, \Actions, Q  \cup
\{\ph,\pf,\ptop,\pbot\}, $ $\delta^0 \wedge \ph,$ $ \nu, \chi' \ra$,
where for every $\sigma \in \Sigma$, $\dashv\,\in \{\bot,\top\}$, the
transition function $\nu$ is defined as follows. Note that the
alphabet of  $\A_\Psi$ is $\Upsilon$, rather than   $\Actions \times
\{\bot,\top\}$. Since, however, base formulas refer to a single
strategy profile and history variable, we restrict attention to the
relevant  components of the input alphabet.
\vspace{-3mm}
\begin{center}
\begin{tabular}{lcl}
$\bullet$
$\nu (\ptop,\la\sigma,\dashv \ra)=\ptop$ and $\nu (\pbot,\la \sigma,\dashv\ra)=\pbot$.
&
&
$\bullet$
$\nu(\ph,\!\la\sigma,\top\ra)=
\bigvee_{d\in \Actions} {((d,\ph) \wedge \bigwedge_{d'\in\Sigma\setminus\{d\}}(d',\ptop))}$.
\\[2mm]
$\bullet$
For every $q\myin Q$, we have
$\nu(q,\!\la \sigma,\dashv \ra )= \delta(q,\sigma)$.
&
&
$\bullet$
$\nu(\pf,\!\la\sigma,\top\ra)=
\bigwedge_{d\in \Actions} (d,\pbot) $.
\\[2mm]
$\bullet$
$\nu(\ph,\!\la\sigma,\bot\ra)=\bigwedge_{d\in \Actions} (d,\pf) $.
&
&
$\bullet$
$\nu(\pf,\!\la\sigma,\bot\ra)=
\bigwedge_{d\in \Actions}{(\bigwedge_{d =\sigma} (d, \pf) \wedge \bigwedge_{d \neq\sigma} (d,\ptop))}$.
\end{tabular}
\\[3mm]
\end{center}

The parity condition $\chi'$ is such that ${\chi'(q) = \chi(q)}$ for every ${q\in Q}$ and for the other states we have  ${\chi'(\ptop)=0}$, ${\chi'(\pbot)=1}$, ${\chi'(\ph)=1}$, and ${\chi'(\pf)=0}$.

It is easy to see that a tree ${\la \Sigma^* ,\tau \ra}$ is accepted by $\A_\Psi$ iff there is a  word ${w\in\Sigma^*}$ such that for every prefix $u$ of $w$ the node $u$ is labeled ${\la \sigma, \top\ra}$ for some ${\sigma\in\Sigma}$ and ${\outcome(\tau)_{w}\models \g}$. The number of states of $\A_\Psi$ is exponential in $\f$ and its index is $3$.


\subsection{Proof of Theorem~\ref{single strategy}}

Consider a lattice $\L$. An element $x\in \L$ is {\em join irreducible\/} if for all
$y,z\in \L$ we have $x \le y\join z$ implies $x\le y$ or $x\le z$.
Given $l$, we define the game $G_l$ as follows.
Let ${X_l = \{ x\in JI(\L)~|~x\le l\}}$ be the set of join irreducible
elements smaller then $l$.
By Birkhoff's representation theorem, a strategy ensures a value
greater or equal $l$ iff for every $x\in X_l$ the strategy ensures a
value greater or equal $x$.

By the analysis in \cite{KL07b}, the value of a latticed play $p$ in a
game $G$
can be decomposed into three values: the acceptance value $acc(p)$, and two values
$r^\join$ and $r^\meet$ that have to do with value relinquished by the
$\join$-player and the $\meet$-player during the play, respectively.
Furthermore, the values $r^\join$ and $r^\meet$ are the limits of the sequences
$\{r^\join_i\}_{i=0}^\infty$ and $\{r^\meet_i\}_{i=0}^\infty$ where
for every $i\ge 0$ the values of $r^\join_i$ and $r^\meet_i$ depend on
the $i$-long prefix of the play $p$.

The idea underlying the reduction is to consider a Boolean game in which
the values from the latticed game are made explicit by the structure of the game
graph. Formally, for a latticed game ${G = \{V, E\}}$ with
${V=V_\join\cup V_\meet}$ and an $\L$-\buchi\  condition $F\in \L^V$, we
define a Boolean generalized-\buchi\  game ${G_l' = \{V',E'\}}$ as follows.
The state space ${V' = V\times \L \times \L}$ is such that in a state
${(u, x, y)\in V\times \L \times \L}$, we have that
$u$ stands for a state in $G$, the value $x$ stands for the
$\join$-relinquished value $r^\join_i$, and the value $y$ stands for the
$\meet$-relinquished value $r^\meet_i$.

Let $G=\{V,E\}$ be a latticed game with an $\L$-\buchi\  condition
$F\in \L^V$ and initial vertex $v_0\in V$.
The {\em simplification} of $G$ for $l\in\L$, denoted $G_l'$, is  the Boolean game $G_l'=\{V',E'\}$ where $V' = V \times\L\times \L$, and the partition of $V'$ and $E'$ is defined as follows.
First,  $V'_\join=V_\join\times\L\times\L$ and
$V'_\meet=V_\meet\times\L\times\L$ (note that even though $G'_l$ is Boolean, we keep the names $\join$-player and $\meet$-player).
The initial vertex is $\tuple{v_0,\top,\bot}$.
In order to define the edges we introduce the following notation.
For $u,u'\in V$ and $x,y\in \L$ the {\em $u'$-successor of $\tuple{u,x,y}$}
is $\tuple{u',x',y'}$, where either $u\in V_\join$ in which case
$x' = x \meet (E(u,v)\join y)$ and $y' = y$, or $u\in V_\meet$ in
which case  $x' = x$ and $y' = y \join (E(u,v)\meet x)$.
Now,
$E' = \{ ( \tuple{u,x,y},$ $\tuple{u',x',y'} ) ~|~
\tuple{u',x',y'}$ is the $u'$-successor of $\tuple{u,x,y}\}$.

It is left to define the generalized-B\"uchi condition.
In order to ensure the value $l\in\L$, the $\join$-player must ``collect'' every
value $x\in X_l$ either as a value relinquished by the $\meet$-player
or by the acceptance value $acc$.
For that, we define, for each $x\in X_l$ a set $F_x$ in the generalized-\buchi\
condition. We define $F_x=(V\times \L \times \{ y\in \L ~|~ y\ge x \}) \cup
(\{ u\in V ~|~ F(u)\ge x) \setminus V\times \{ y\in \L ~|~ y\not \ge x \}
\times \L)$. The first component states for states in which
the $\meet$-player relinquished $x$, and the second component stands for states in  which both the acceptance value is greater then $x$ and $x$ was
not relinquished by the $\join$-player in the past.
Now, the generalized-\buchi\  acceptance condition is
$F' = \{ F_x ~|~ x\in X_l \} $.

Assume first there exists a single strategy $\pi$ in $G$ ensuring value
greater or equal $l$.
Every strategy $\pi$ for $G$ (for either player) induces a strategy $\pi'$
in $G_l'$ in which
$\pi'( \tuple{u_0,x_0,y_0}, \ldots, \tuple{u_n,x_n,y_n} )$ is the
$\pi(u_0, \ldots, u_n)$-successor of $\tuple{u_n,x_n,y_n}$.
Consider a $\join$-player strategy $\pi$ that ensures value greater or equal
$l$. We show that $\pi'$ is winning in $G_l'$.
It is not hard to see that a play
$p' = \tuple{u_0,x_0,y_0} \ldots \tuple{u_n,x_n,y_n} \ldots$
consistent with $\pi'$ corresponds to a play
$p = u_0\ldots u_n \ldots$ consistent with $\pi$.
Furthermore, for every $i\ge 0$, we have $x_i = r^\join_i$ and $y_i = r^\meet_i$.
Since $\pi$ ensures value $l$ in $G$, the value of $p$ is greater or
equal $l$, and therefore, for every join irreducible $x\in V_x$ we
have $val(p) \ge x$. Thus, either there exists an index $i$ from which
$r^\meet_i\le x$ or for infinitely many $i$'s we have $F(u_i)\ge x$ and
$r^\join_i\ge x$. Both cases imply that the set $F_x$ is traversed
infinitely often. Thus the play $p'$ is winning for the $\join$-player in $G_l'$.

Assume now that $\pi'$ is a winning strategy for the $\join$-player in $G_l'$.
The strategy $\pi'$ induces a $\join$-player strategy in $G$ in the
following way: Every prefix of a play $p = u_0,u_1,\ldots,u_n$ in $G$ induces
the prefix of a play
$p' = \tuple{u_0,\top,\bot},\tuple{u_0,x_1,y_1},  \ldots, \tuple{u_n,x_n,y_n}$,
where for every $i>0$, we have that $\tuple{u_i,x_i,y_i}$ is the
 $u_i$-successor of $\tuple{u_{i-1},x_{i-1},y_{i-1}}$.
We define $\pi(p)$ to be the state $u$ for which $\pi'(p')$ is $\tuple{u,x,y}$.
It is not hard to see that for a play $p$ in $G$ consistent
with $\pi$, and for every $i\ge 0$, we have $x_i = r^\join_i$ and $y_i =
r^\meet_i$.
As $\pi'$ is winning in $G_l'$, we get that for every $x\in X_l$ we have
$val(p)\ge x$, and therefore $val(p)\ge l$.

\subsection{Proof of Theorem~\ref{thm:lnashge}}

Approaching the problem in a fashion similar to the one we used in the
Boolean case, we represent  strategy profiles by
$\Sigma$-labeled $\Sigma$-trees, and sets of profiles by tree
automata.
We construct two Boolean tree automata.
The first, denoted $\A_0$, for the language of all profiles $\pi$ in which
$\payoff_0(\pi)\ge v$, and the second, denoted $\A_N$, for the
language of all Nash equilibria. It is not hard to see that the
intersection of $\A_0$ and $\A_N$ contains all the solutions to the
latticed rational synthesis problem.
Thus, solving the problem amounts to returning a witness to the nonemptiness of the intersection.

For the purposes of complexity analysis, we denote by $s_i$ the size of
the LDBW for the $i$-th agent specification, by $s = max \{ s_i\}$ the
maximal $s_i$, and by $m = |\L|$ the
size of the lattice.

We first construct $\A_0$.
As in the Boolean case, we first construct an
LDBT $\A'_0$ that maps a strategy profile $\pi$ to $\payoff_0(\pi)$.
Using Theorem~\ref{single strategy}, we can
construct from $\A'_0$ the required Boolean tree automaton $\A_0$.
To see how, note that the generalized-\buchi\  game involved has a very uniform
structure. From every $\join$-vertex, the $\join$-player has exactly one choice
associated with each $\sigma\in \Sigma$. (This property is inherited
from the latticed game which in turn inherits it from the fact that the alphabet of
$\A'_0$ is $\Sigma$.)
A similar property holds for the $\meet$-player (this property is inherited from the fact that $\A'_0$
runs on $\Sigma$-trees). Therefore, the generalized-\buchi\  game can
be reduced, using standard techniques, to a generalized-\buchi\ tree automaton $\A_0$.
The size of $\A'_0$ is $s_0\cdot m^2$ and the number of acceptance
sets in its generalized \buchi\ condition is bounded by $m$.

We now turn to build an automaton for Nash equilibria $\A_N$.
We construct $\A_N$ as an intersection of $n$ automata
$\{ \A_N^i\}_{i=1}^n$, where the language of $\A_N^i$ is the set of the profiles that
satisfy $\payoff_i(\pi_{-i},\pi'_i) \le \payoff_i(\pi)$.
By Birkhoff's representation theorem, an equivalent criteria would be that for
every  join irreducible element $j\in JI(\L)$, we have
$\payoff_i(\pi_{-i},\pi'_i) \ge j \rightarrow  \payoff_i(\pi,\varphi_i)\ge j$.
Given LDBW for $\varphi_i$, it is not hard to construct LDBTs for
$\payoff_i(\pi_{-i},\pi'_i)$ and  $\payoff_i(\pi)$.
For every join irreducible element $j\in JI(\L)$ we would like to make sure
that $\payoff_i(\pi_{-i},\pi'_i) \ge j \rightarrow \payoff_i(\pi,\varphi_i)\ge j$.
To that end, we use the construction of the Boolean game $\G_\top$ in the proof of
Theorem~\ref{single strategy}. Recall that in the game $\G_\top$, the value $x$ is
obtained by a single strategy iff the acceptance set $F_x$ is visited
infinitely often. Thus, for a specific agent $i\le n$, and a join
irreducible element $j\in JI(L)$, we can construct a Boolean \buchi\  tree
automaton $\B^i_j$, of size $O(s_i\cdot m^2)$,  that accepts exactly the
trees encoding profiles for which $\payoff_i(\pi,\varphi_i)\ge
j$. In a similar way, we can construct a tree automaton $C^i_j$, of similar
size, that accepts trees encoding profiles for which
$\payoff_i(\pi_{-i},\pi'_i) \ge j$. Combining $B^i_j$ and $C^i_j$ we
can get a Streett automaton $A^i_j$ that accepts profiles for which
$\payoff_i(\pi_{-i},\pi'_i) \ge j \rightarrow \payoff_i(\pi,\varphi_i)\ge j$.
The size of $A^i_j$ is $O(s_i^2\times m^4)$, and it has one Streett pair.
Note that for a fixed $i$, the automata $A^i_j$ share their structure and
only differ in the acceptance condition. Therefore, for a fixed $i\le n$, we
can construct an automaton $A^i_N$, of size $O(s_i^2\cdot m^4)$ and with
$O(m)$ pairs, that accepts profiles for which
$\payoff_i(\pi_{-i},\pi'_i) \ge j \rightarrow
\payoff_i(\pi,\varphi_i)\ge j$ for every join irreducible element $j\in JI(\L)$.
By intersecting the automata $\A^i_N$ we get an automaton $\A_N$ of size
$(s\cdot m)^{O(n)}$, with $O(m\cdot n)$ pairs.

The intersection of $\A_0$ and $\A_N$ is a Streett automaton
of size $(s\cdot m)^{O(n)}$ and with $O(m\cdot n)$ pairs. Its
emptiness can then be checked in time $(s\cdot m)^{O(m\cdot n^2)}$
\cite{KV98}, and we are done.


\begin{thebibliography}{10}

\bibitem{AKL09}
B.~Aminof, O.~Kupferman, and R.~Lampert.
\newblock Reasoning about online algorithms with weighted automata.
\newblock In {\em Proc.\ 20th SODA}, pages
  835--844, 2009.

\bibitem{AAE04}
P.C. Attie, A.~Arora, and E.A. Emerson.
\newblock Synthesis of fault-tolerant concurrent programs.
\newblock {\em TOPLAS},
  26:128--185, 2004.

\bibitem{CCHKM05}
A.~Chakrabarti, K.~Chatterjee, T.A. Henzinger, O.~Kupferman, and R.~Majumdar.
\newblock Verifying quantitative properties using bound functions.
\newblock In {\em Proc. 13th CHARME}, LNCS 3725, pages  50--64, 2005.

\bibitem{CDH08}
K.~Chatterjee, L.~Doyen, and T.~Henzinger.
\newblock Quantative languages.
\newblock In {\em Proc. 17th CSL}, LNCS 5213, pages 385-400,  2008.

\bibitem{CHJ08}
K.~Chatterjee, T.~Henzinger, and B.~Jobstmann.
\newblock Environment assumptions for synthesis.
\newblock In {\em Proc. 19th CONCUR}, LNCS 5201, pages 147--161, 2008.

\bibitem{CHJ06}
K.~Chatterjee, T.~Henzinger, and M.~Jurdzinski.
\newblock Games with secure equilibria.
\newblock {\em Theoretical Computer Science}, 2006.

\bibitem{CHP07}
K.~Chatterjee, T.~A. Henzinger, and N.~Piterman.
\newblock Strategy logic.
\newblock In {\em 18th CONCUR}, LNCS, pages 59--73, 2007.

\bibitem{CH07}
K.~Chatterjee and T.A. Henzinger.
\newblock Assume-guarantee synthesis.
\newblock In {\em Proc.\ 13th TACAS}, LNCS 4424, pages 261--275, 2007.

\bibitem{CMJ04}
K.~Chatterjee, R.~Majumdar, and M.~Jurdzinski.
\newblock On {Nash} equilibria in stochastic games.
\newblock In {\em Proc. 13th CSL}, LNCS 3210, pages 26--40, 2004.

\bibitem{Chu63}
A.~Church.
\newblock Logic, arithmetics, and automata.
\newblock In {\em Proc. Int. Congress of Mathematicians, 1962}, pages 23--35, 1963.

\bibitem{GTW02}
E.~Gr{\"a}del, W.~Thomas, and T.~Wilke.
\newblock {\em Automata, Logics, and Infinite Games: A Guide to Current
  Research}, LNCS 2500, 2002.

\bibitem{GH82}
Y.~Gurevich and L.~Harrington.
\newblock Trees, automata, and games.
\newblock In {\em Proc.\ 14th STOC}, pages 60--65, 1982.

\bibitem{GC03}
A.~Gurfinkel and M.~Chechik.
\newblock Multi-valued model-checking via classical model-checking.
\newblock In {\em 14th CONCUR}, LNCS 2761, pages 263--277, 2003.

\bibitem{JGB05}
B.~Jobstmann, A.~Griesmayer, and R.~Bloem.
\newblock Program repair as a game.
\newblock In {\em Proc 17th CAV}, LNCS
  3576, pages 226--238, 2005.

\bibitem{KL07}
O.~Kupferman and Y.~Lustig.
\newblock Lattice automata.
\newblock In {\em Proc. 8th VMCAI}, LNCS 4349, pages 199 -- 213, 2007.

\bibitem{KL07b}
O.~Kupferman and Y.~Lustig.
\newblock Latticed simulation relations and games.
\newblock In {\em 5th ATVA}, LNCS 4762, pages
  316--330, 2007.

\bibitem{KPV06}
O.~Kupferman, N.~Piterman, and M.Y. Vardi.
\newblock Safraless compositional synthesis.
\newblock In {\em Proc 18th CAV}, LNCS 4144, pages 31--44,
  2006.

\bibitem{KV01}
O.~Kupferman and M.Y. Vardi.
\newblock Synthesizing distributed systems.
\newblock In {\em Proc.\ 16th LICS}, pages
  389--398, 2001.

\bibitem{KV05a}
O.~Kupferman and M.Y. Vardi.
\newblock Safraless decision procedures,
\newblock In {\em Proc.\ 46th FOCS}, pages 531--540, 2005.

\bibitem{KV05b}
O.~Kupferman and M.Y. Vardi.
\newblock From linear time to branching time.
\newblock {\em TOCL}, 6(2):273--294, 2005.

\bibitem{KV06a}
O.~Kupferman and M.Y. Vardi.
\newblock Memoryful branching-time logics.
\newblock In {\em Proc.\ 21st LICS}, pages  265--274, 2006.

\bibitem{KV98}
O.~Kupferman and M.Y. Vardi.
\newblock Weak alternating automata and tree automata emptiness.
\newblock In {\em Proc.\ 30th STOC}, pages 224--233, 1998.

\bibitem{LV09}
Y.~Lustig and M.Y. Vardi.
\newblock Synthesis from component libraries.
\newblock In {\em Proc. 12th FOSSACS}, LNCS 5504, pages 395--409, 2009.

\bibitem{MP92}
Z.~Manna and A.~Pnueli.
\newblock {\em The Temporal Logic of Reactive and Concurrent Systems:
  Specification}.
\newblock Springer, 1992.

\bibitem{MW05}
R.~{van der} Meyden and T.~Wilke.
\newblock Synthesis of distributed systems from knowledge-based specifications.
\newblock In {\em 16th CONCUR}, LNCS 3653, pages 562--576, 2005.

\bibitem{MS87}
D.E. Muller and P.E. Schupp.
\newblock Alternating automata on infinite trees.
\newblock {\em Theoretical Computer Science}, 54:267--276, 1987.

\bibitem{Nas50}
J.F. Nash.
\newblock Equilibrium points in n-person games.
\newblock In {\em Proceedings of the National Academy of Sciences of the United
  States of America}, 1950.

\bibitem{NR99}
N.~Nisan and A.~Ronen.
\newblock Algorithmic mechanism design.
\newblock In {\em Proc.\ 31st STOC}, pages
  129--140, 1999.

\bibitem{NRTV07}
N.~Nisan, T.~Roughgarden, E.~Tardos, and V.~V. Vazirani.
\newblock {\em Algorithmic Game Theory}.
\newblock Cambridge University Press, 2007.

\bibitem{OR94}
M.~J. Osborne and A.~Rubinstein.
\newblock {\em A Course in Game Theory}.
\newblock The MIT Press, 1994.

\bibitem{Pnu77}
A.~Pnueli.
\newblock The temporal logic of programs.
\newblock In {\em Proc.\ 18th FOCS},
  pages 46--57, 1977.

\bibitem{PR89a}
A.~Pnueli and R.~Rosner.
\newblock On the synthesis of a reactive module.
\newblock In {\em Proc.\ 16th POPL}, pages 179--190, 1989.

\bibitem{RW89}
P.J.G. Ramadge and W.M. Wonham.
\newblock The control of discrete event systems.
\newblock {\em {IEEE} Transactions on Control Theory}, 77:81--98, 1989.

\bibitem{Ros92}
R.~Rosner.
\newblock {\em Modular Synthesis of Reactive Systems}.
\newblock PhD thesis, Weizmann Institute of Science, 1992.

\bibitem{Sel75}
R.~Selten.
\newblock Reexamination of the perfectness concept for equilibrium points in
  extensive games.
\newblock {\em International Journal of Game Theory}, 4(1):25--55, March 1975.

\bibitem{SG05}
S.~Shoham and O.~Grumberg.
\newblock Multi-valued model checking games.
\newblock In {\em Proc. 3rd ATVA}, LNCS 3707, pages 354--369, 2005.

\bibitem{Umm06}
M.~Ummels.
\newblock Rational behaviour and strategy construction in infinite multiplayer
  games.
\newblock In {\em Proc. 26th FSTTCS}, LNCS 4337, pages 212--223, 2006.

\bibitem{VW94}
M.Y. Vardi and P.~Wolper.
\newblock Reasoning about infinite computations.
\newblock {\em Information and Computation}, 115(1):1--37, 1994.

\end{thebibliography}
\end{document}